\documentclass[12pt]{article}
\usepackage{setspace,graphicx,rotating,epsfig}
\usepackage{amsmath,amssymb,amsfonts}
\usepackage{chicago}
\input{epsf}
\newcommand{\bvec}[1]{\mbox{\boldmath $#1$}}
\newcommand{\btens}[1]{\mbox{\boldmath $#1$}}

\newcommand{\tauyx}{\mbox{$\tau\!_{yx}$}}
\newcommand{\tauyy}{\mbox{$\tau\!_{yy}$}}

\newcommand{\tauzy}{\mbox{$\tau\!_{zy}$}}

\newcommand{\tauxxb}{\mbox{$\bar{\tau}\!_{xx}$}}
\newcommand{\tauyxb}{\mbox{$\bar{\tau}\!_{yx}$}}
\newcommand{\tauyyb}{\mbox{$\bar{\tau}\!_{yy}$}}
\newcommand{\tauzxb}{\mbox{$\bar{\tau}\!_{zx}$}}
\newcommand{\tauzyb}{\mbox{$\bar{\tau}\!_{zy}$}}
\newcommand{\tauzzb}{\mbox{$\bar{\tau}\!_{zz}$}}

\newcommand{\gdot}{\mbox{$\dot{\gamma}$}}
\newcommand{\we}{W\!\!e}

\newcommand{\bnabla}{\boldsymbol{\nabla}}
\begin{document}
    \begin{spacing}{1}
\title{Slip, concentration fluctuations and flow instability in sheared
polymer solutions}
\author{William B. Black\footnote{Present address: 3M Corporate
        Process Technology Center, St. Paul, MN, 55144} and Michael
        D. Graham\footnote{Corresponding author.  E-mail:
        graham@engr.wisc.edu}\\ Department of Chemical Engineering and Rheology Research
         Center \\ University of Wisconsin-Madison,  Madison, WI
         53706-1691}
% \address{}
\date{\today}
\maketitle

\begin{abstract}
Recent experiments suggest that shear-enhanced growth of concentration
fluctuations in polymer solutions is strongly influenced by solid
boundaries.  We analyze the dynamics of a model of a sheared polymer
solution, accounting for the effect of stress variations on mass flux
and for wall-slip.  If (and only if) these effects are present, the
flow exhibits a boundary-localized instability consistent with
experimental observations.  Even in the absence of flow instability,
thermally-driven concentration fluctuations are significantly enhanced
near the boundary.

\end{abstract}

% \pacs{PACS numbers:61.25.Hq, 83.50.Lh}

The local composition of polymer solutions and blends can be
significantly affected by flow.  In particular, shear-enhanced
concentration fluctuations in semidilute polystyrene (PS) solutions
have been observed via light-scattering and microscopy by a number of
groups
\cite{fujioka1991,kume1992,dixon1991,wu1992,hashimoto1994,wirtz1994a,mizukami1996,fuller1997}.
A purely hydrodynamic mechanism for this phenomenon, in an unbounded
flow domain, was proposed initially by Helfand and Fredrickson
(HF)~\cite{fredrickson1989} and later extended by several
authors~\cite{milner1991,DoiOnuki92,milner1993,helfand1995,jasnow1997}.  This
mechanism is based on the idea that variations in polymer stress can
drive a flux of polymer molecules -- indeed, several
authors~\cite{Onuki90,milner1993,helfand1995,beris1992} have developed simple,
two-fluid models which demonstrate quite generally that the polymer
flux in a solution, $\bvec{j}$, has the general form $\bvec{j} = -D \bnabla n +
(D/k_{B} T) \bnabla \cdot \btens{\tau}$, where $n$ is the polymer
number density, $\btens{\tau}$ is the polymer extra stress tensor, $D$
is the polymer gradient diffusivity, $k_{B}$ is the Boltzmann
constant, and $T$ is the temperature.  Furthermore, a detailed kinetic
theory derivation for a dilute solution~\cite{rbbird1999} leads to a very similar expression
for the flux.  A scattering peak is predicted to arise when the
diffusion and relaxation times are equal, which occurs at a wavelength
of $2 \pi/k \approx \sqrt{D \lambda}$ which is, not coincidentally,
the only inherent length scale in the problem.

\begin{figure}
%   \centerline{\epsfig{figure=geometry3.eps,width=2.5in}}
  \centerline{\epsfig{figure=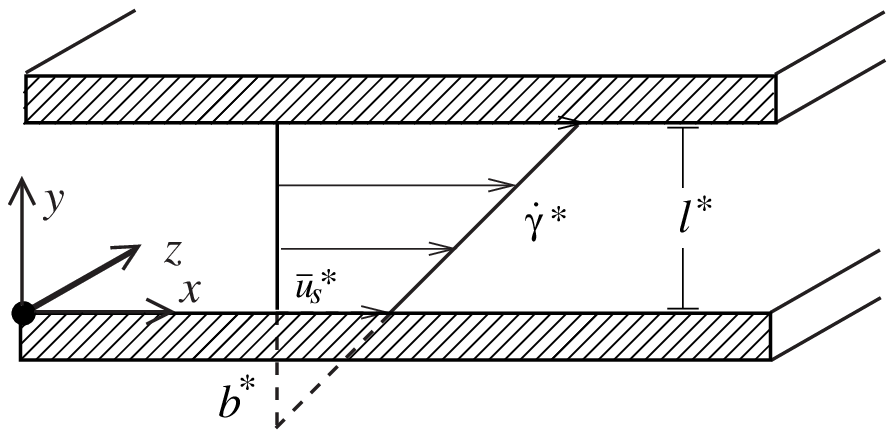,width=2.5in}}
   \caption{Plane Couette geometry showing slip between the solution
   and solid surface.  The flow is in the $x$-direction, $y$ is the
   gradient direction and $z$ is the vorticity or neutral direction.
   $l^*$ is the gap width, $u_s^*$ is the slip velocity,
   $\dot{\gamma}^*$ is the true shear rate, and $b^* \equiv
   u_s^*/\dot{\gamma}^*$ is the steady-state extrapolation length.  The
   asterisks (here and in the text) denote dimensional quantities.}
   \label{geometry}
\end{figure}

This length scale is observed in light scattering
experiments.  Hashimoto and
coworkers~\cite{fujioka1991,kume1992,hashimoto1994,fuller1997} have
performed a series of experiments with PS dissolved in
dioctylphthalate (DOP) and have measured the scattering intensity by
illuminating along the shear-gradient direction (the $y$-direction of
Fig.~\ref{geometry}).  They observed a ``butterfly'' scattering
pattern indicative of flow structures with wave vectors roughly
aligned in the flow direction.  The wavenumbers obtained from the
patterns were on the order of $k \approx 0.6 \mu$m$^{-1}$ giving a
length scale for the fluctuations of approximately $10 \mu$m.
Estimating the relaxation time as the reciprocal of the shear rate for
the onset of shear thinning~\cite{fuller1997} and using the
diffusivity measured by Wu \textit{et al.}~\cite{dixon1991} for a
similar system gives a predicted wavenumber of $k \approx 0.7
\mu$m$^{-1}$, which is close to the measured value.  This length scale
was also observed by Wu \textit{et al.}~\cite{dixon1991}, who measured
the scattering intensity by illuminating along the vorticity direction
(the $z$-direction in Fig.~\ref{geometry}).  At low shear rates, the
scattering peak was located at $k \approx 10 \mu$m$^{-1}$, which is in
good agreement with their predicted value of $9.4 \mu$m$^{-1}$.  

Even though both of these experiments show peaks at the requisite
length scale, qualitative differences exist in the shear rate
dependence of the results.  In particular, Hashimoto and coworkers,
who measured the intensity essentially averaged over a volume
containing both near-surface regions, found that the intensity
increased markedly above a critical shear rate, which was roughly the
reciprocal of the polymer relaxation time (an approximately 100-fold
increase in scattering for a 10-fold increase in shear rate).
Conversely, Wu \textit{et al.}, whose measurements sample only the
bulk, not the near-surface regions, found a much smaller increase in
intensity as the shear rate was increased, approximately an 11-fold
increase in scattering intensity for a 25-fold increase in shear rate.

A recent experimental study by Mhetar and Archer~\cite{archer1998} has
demonstrated several novel features of shear enhanced concentration
fluctuations in entangled polymer solutions.  They studied semidilute
solutions of PS in diethylphthalate (DEP), to which tracer particles
were added to measure the velocity profile, using a planar Couette
cell %\emph{
with (high energy) silica walls.  In this situation,
wall-slip occurs primarily due to the disentanglement of bulk chains
from those adsorbed to the walls, rather than to the formation of a 
depletion layer, as arises in dilute solutions 
\cite{Graham99,Aussere82}.  Two features of the flow of these
solutions stand out.  First, no turbidity was observed at low shear 
stresses where slip was weak;  
enhanced concentration fluctuations were observed, however, at large
shear stresses, where slip was significant.  The length scale for the fluctuations was
observed to be $\approx 10 \mu$m, as was the extrapolation, or
slip, length measured for these solutions.  %}  
Second, the concentration
fluctuation enhancement began near the boundaries and modification of
the surface to increase slip delayed the onset on fluctuations to much
larger shear stresses.  These observations strongly suggest that slip
plays a role in the formation and development of these fluctuations.

Taken together, experiments suggest that the HF mechanism, while being
able to explain bulk behavior, only partially accounts for the physics
leading to enhancement of fluctuations near surfaces, a conclusion
that can be put into perspective by consideration of the fluid
dynamics of polymer flows near boundaries.  Gorodtsov and
Leonov~\cite{leonov1967} analyzed the plane Couette flow of an upper
convected Maxwell (UCM) fluid (a model of a melt) and found that,
while the flow is always linearly stable, the most slowly decaying
modes are localized near the bounding surfaces.  As a result, we might
expect the formation of boundary localized structures due to random
concentration fluctuations in the polymer solution.  In addition, slip
has been shown to lead to boundary-localized flow instability in
polymeric simple shear flows under certain
conditions~\cite{mrenardy1990,graham1996,graham1999b}.  This suggests
that boundary-localized concentration fluctuation dynamics can appear
by analogous routes.  Our goal here is to analyze the coupling between
concentration, stress, and slip in bounded shear flow, to elucidate
the effects of boundaries on the dynamics.

We consider flow in the plane Couette geometry shown in
Fig.~\ref{geometry}.  
% To describe the behavior of the solution, we use
% the two-fluid model derived by Mavrantzas and Beris~\cite{beris1992},
% simplified to a solution of Hookean
% dumbbells~\cite{mavrantzas1994}, 
To describe the behavior of the solution, we use
the kinetic theory model derived by Beris and Mavrantzas~\cite{mavrantzas1994},
for a solution of Hookean
dumbbells, 
and written in terms of the
conformation tensor, $\btens{\alpha} = \langle
\bvec{Q}\bvec{Q}\rangle$, which is the second moment of the
orientation vector of the dumbbells, $\bvec{Q}$.  Nondimensionalizing
stresses with $n_0^* k_{B}T$, where $n_0^*$ is the average concentration
of the solution, lengths with the characteristic length $\sqrt{D 
\lambda}$,
% (with gradient diffusivity $D$ evaluated at the nominal 
% concentration $n_{0}^{*}$), 
 velocities with $\gdot^* \sqrt{D \lambda}$, time
with $\lambda$, and concentration with $n_0^*$ gives the following
model equations:
\begin{eqnarray}
   \frac{\partial \btens{\alpha}}{\partial t} +
   \bvec{v}\cdot \bnabla \btens{\alpha} - \btens{\alpha}  
    \cdot \bnabla \bvec{v} - (\bnabla \bvec{v})^T \cdot
   \btens{\alpha}  + 
   \btens{\alpha} - n \btens{\delta}= \nabla^{2}\btens{\alpha}
   \label{nondimconst}, \\
   \frac{\partial n}{\partial t} + \bvec{v}\cdot \bnabla n =
   - \bnabla \cdot \left(-\boldsymbol{\nabla} n + \bnabla
   \cdot \btens{\tau} \right), \label{nondimconcentration} \\
   -\bnabla p + \bnabla \cdot \btens{\tau} + S \, \bnabla^2 \bvec{v}
   = 0,  \label{nondimmomentum} \\
   \bnabla \cdot \bvec{v} = 0, \label{nondimcontinuity}
\end{eqnarray}
where $\bvec{v}$ is the velocity, $S$ is the ratio of the solvent to
polymer viscosities, $n$ is the number density of dumbbells,
$\btens{\tau} = \btens{\alpha} - n \btens{\delta}$ is the polymer
extra stress and $p$ is the pressure.  In presenting the model, we
have suppressed any dependencies of properties on concentration -- in
the analysis below we only consider small perturbations from spatially
uniform stress and concentration fields, in which case the terms that
reflect these dependencies do not appear.  Eq.~\ref{nondimconst}
describes the evolution of polymer conformation and is a
generalization of the upper convected Maxwell (or Oldroyd-B)
model~\cite{rbbirdb} to include spatial diffusion of conformation (on
the RHS).  In most cases, this term is
negligible~\cite{mavrantzas1994}, but here, the flow structures we are
interested in have a length scale set by the diffusivity. 
Eq.~\ref{nondimconcentration} is the conservation equation for the
dumbbells, explicitly showing the dependence of the polymer flux on
stress variations.  If diffusion of the polymer molecules is
neglected, the right-hand sides of Eqs.~\ref{nondimconst} and
\ref{nondimconcentration} vanish, recovering the Maxwell model. 
Eqs.~\ref{nondimmomentum} and \ref{nondimcontinuity} describe
conservation of momentum and mass.  Other phenomenologically motivated
models for such systems are virtually identical to this one, but do
not generally include the conformational diffusion
term~\cite{fredrickson1989,milner1991,milner1993,helfand1995,jasnow1997},
the primary difference being the use of the polymer phase velocity 
rather than the mass-average velocity in
the deformation terms of the equation for polymer conformation -- our
preliminary results with such a full two-fluid model yield
qualitatively identical results to those presented below.
% Previously, however, only Sun \textit{et
% al.}~\cite{jasnow1997} have included the diffusive term in
% Eq.~\ref{nondimconst} in a treatment of scattering.
Furthermore, all
earlier works have restricted themselves to bulk behavior and $\we <
1$.

% \emph{
Because of the high energy surfaces in Mhetar and Archer's
experiments, it is reasonable to assume that the adsorbed chains do
not desorb to a significant degree in flow, and that no additional
bulk chains adsorb.  Therefore, we apply a no-flux boundary condition
for the polymer concentration: $\bvec{n}\cdot\bvec{j} = \partial
n/\partial y - \partial \tauyx /\partial x - \partial \tauyy/\partial
y - \partial \tauzy/\partial z$ = 0, where $\bvec{n}$ is the unit
normal on the boundary.  Inclusion of the diffusion term in the
constitutive equation necessitates the specification of boundary
conditions for the conformation tensor.  For the same reason as above,
we assume that there are no sources or sinks for polymer conformation
at the walls, giving a no-flux boundary condition for the
conformation, $\partial \btens{\alpha}/\partial y=0$.  We have also
performed computations using the classical UCM equation as the
boundary condition for $\btens{\alpha}$~\cite{beris1995b} and found no
qualitative difference in the results.
% }

The model is completed by specifying the slip boundary condition at
the interface, which crudely reflects the interaction between the 
bulk chains and those attached to the wall.  We use here the simplest nontrivial model, the Navier
slip boundary condition $u_s = \epsilon \tauyx (= b =
b^*/\sqrt{D\lambda} \mbox{ at steady state})$; $\epsilon$ is given
by $b/\we$.

We have studied the dynamics of this model primarily
by linear stability analysis of the plane Couette base state, given by
$\bvec{\bar{a}} = (\bar{v}_x , \bar{v}_y , \bar{v}_z , \tauxxb ,
\tauyxb , \tauzxb , \tauyyb , \tauzyb , \tauzzb , \bar{n},\bar{p}) \\
\nonumber = ( \we(y + b) , 0 , 0 , 2 \we^2 , \we , 0 , 0 , 0 , 0 , 1, 1)$,
where $\btens{\alpha}$ has been eliminated in favor
of $\btens{\tau}$.  Small perturbations are added to this base
solution, i.~e.~$\bvec{a} = \bar{\bvec{a}}(y) +
\tilde{\bvec{a}}(x,y,z,t)$, and only the terms linear in the
perturbation are retained in the model equations.  The perturbations
take the normal mode form: $\tilde{\bvec{a}}(x,y,z,t) =
\hat{\bvec{a}}(y) e^{i\boldsymbol{k}\cdot\boldsymbol{x}} e^{-i \sigma
t} + \mbox{c.c.}$, where $\bvec{k} = (k_x,0,k_z)$.  This formulation
yields a generalized eigenvalue problem for the eigenvalues
$\{\sigma\}$ with associated eigenfunctions $\{\hat{\bvec{a}}(y)\}$.
If $\mbox{Im}(\sigma) > 0$ then infinitesimal disturbances grow and
the flow is unstable.

Prior work~\cite{graham1996,graham1999b} with the UCM model and
preliminary work with the present model have shown that the
eigenfunctions of the discrete spectrum are localized near the
boundaries.  Furthermore, in experiments, $\sqrt{D\lambda}\ll l^{*}$.
Therefore, we solve the eigenvalue problem on a semi-infinite domain
(SID).  We map $y \in [0,\infty)$ onto the computational domain, $\xi
\in [-1,1]$, using the mapping $y = 0.1 (1 + \xi)/(1 - \xi)$, and then
perform Chebyshev collocation on the mapped system~\cite{zang}, using
$N+1$ collocation points.  The eigenvalue problem is solved using a
public domain routine~\cite{garbow1978}.

\begin{figure}[h]
%   \centerline{\epsfig{figure=neutral.eps,width=2.5in}}
  \centerline{\epsfig{figure=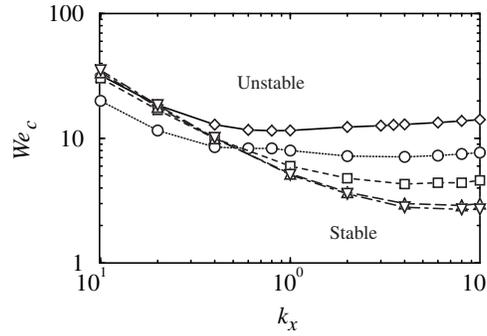,width=2.5in}}
   \caption{Neutral curves for $k_z = 0$, $S = 10^{-2}$, $N=96$,
   and various values of the extrapolation length.  {\large \boldmath
   $\diamond$} - $b =
   1$, {\boldmath $\bigcirc$} - $b = 2$, {\large \boldmath $\square$} -
   $b = 4$, {\boldmath $\bigtriangleup$} - $b =
   8$, and {\boldmath $\bigtriangledown$} - $b = 10$.}
   \label{neutrala}
\end{figure}

To begin the discussion of the results, we note that if the
possibility of concentration fluctuations is suppressed in the model
(by setting $D$ to zero, thus eliminating the RHS of eqs.~\ref{nondimconst} and
\ref{nondimconcentration}), plane shear flow with either the
no-slip~\cite{leonov1967} or Navier
slip~\cite{petrie,graham1996,graham1999b} boundary condition is stable
for all Weissenberg numbers.  In addition, we have found that if
concentration variations are permitted but slip is not (by setting
$b=0$), the flow is also stable.  In contrast, if slip and
concentration variations are allowed, the flow can become unstable
when $\we$ exceeds a critical value $\we_{c}$ -- a new class of
polymer flow instabilities arises.  Fig.~\ref{neutrala} shows neutral
curves, plotted as $\we_c$ vs. $k_x$ for two dimensional disturbances,
i.e. $k_z = 0$, with $b$ fixed.  For $b = O(1)$, the critical
wavenumber is $O(1)$, so that the critical length scale for the
instability is $\sqrt{D \lambda}$.  We have also examined stability
with respect to three dimensional disturbances ($k_z \neq 0$), but
find that two dimensional disturbances are always the most dangerous.
At low wavenumbers (those most easily observed by microscopy),
$\we_{c}$ first decreases and then increases with increasing $b$ --
the increasing section of the curve is in agreement with the
observations of Mhetar and Archer~\cite{archer1998} that treating the
surface to increase slip delayed the onset of fluctuations.  Note that
there is a transition from one mode of instability to another, leading
to multiple minima in the neutral curve, as most clearly seen in the
curve for $b=2$.  For larger values of $b$, the neutral curves
collapse toward a single curve.  Similar results are found for larger
values of $S$, but the critical Weissenberg numbers are much lower,
$\we_c \approx 2$ for $S \gtrsim 1$. (However, for $S = O(1)$, the
solution would typically be too dilute for slip to be a realistic
boundary condition.)

\begin{figure}[h]
%   \centerline{\epsfig{figure=n_nocontours.eps,width=2.5in}}
  \centerline{\epsfig{figure=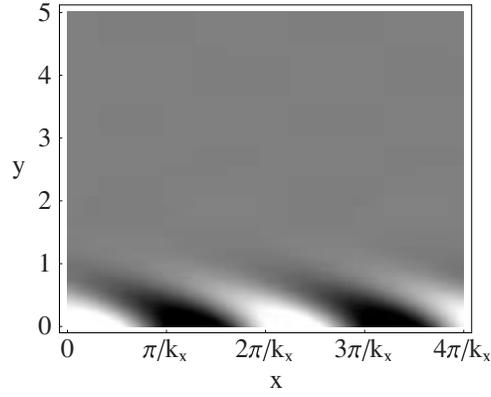,width=2.5in}}
   \caption{Concentration component of the unstable eigenfunction;
   $k_x = 0.4$, $b = 10$, $\we = 10$, $S = 0.01$, $N = 96$.}
   \label{evector}
\end{figure}

Fig.~\ref{evector} shows a density plot of a typical destabilizing
concentration perturbation.  Near the boundary, the local wavevector
has the same orientation as that predicted by the HF
mechanism~\cite{fredrickson1989,helfand1995}.  In agreement with the
observations of Mhetar and Archer~\cite{archer1998}, perturbations are
localized near the surfaces and are observed at large $\we$.  As a
result, we would expect scattering to be more pronounced in the
boundary regions than in the bulk.

We have also investigated the dynamics of Brownian fluctuations in
this model.  Ji and Helfand~\cite{helfand1995} have demonstrated that
the velocity is a fast variable and therefore, since introduction of
concentration fluctuations into this system results in adding
fluctuations directly to the stress, we only incorporate the random
forcing into the polymer mass balance.  This is carried out by adding
a term $\nabla \cdot \bvec{w}$ to Eq.~\ref{nondimconcentration}, where
$\bvec{w}$ is the random contribution to the polymer flux and
satisfies $\langle \bvec{w}(\bvec{x},t) \bvec{w}(\bvec{x}',t')\rangle
= [(2 n_0^* (D\lambda)^{3/2})] \btens{\delta} \, \delta(\bvec{x} -
\bvec{x}') \delta(t - t')$~\cite{gardiner}.  To conserve mass, we
require $w_y = 0$ on $y=0,1$.  We follow the common
procedure~\cite{fredrickson1989,milner1991,milner1993,helfand1995} of
assuming the noise is small and contributes only to the linearized
equations.  Again, Chebyshev collocation is used for the spatial
discretization in the $y$-direction with a Fourier transform in the
$x$-direction, but because the collocation points are not uniformly
distributed, the random forcing is weighted so that it is uniform in
magnitude throughout the domain.  The resulting system of equations
was time-integrated on a bounded domain using an implicit Euler scheme
with the top plate located at $l^* = 100 \sqrt{D \lambda}$.  Results
are collected only after the initial transient has decayed and
fluctuation effects dominate.

\begin{figure}
%   \centerline{\epsfig{figure=conc_noise.eps,width=2.5in}}
  \centerline{\epsfig{figure=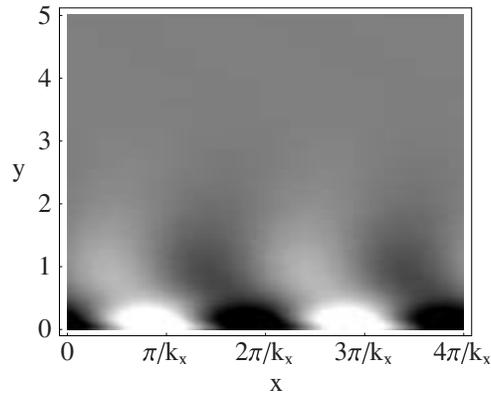,width=2.5in}}
   \caption{Typical snapshot of a concentration profile driven by
            random fluctuations.  The parameters are: $\we = 1$, $k_x
            = 1$, $b = 1$, $S=0.01$, $N = 192$.}
\label{conc_noise}
\end{figure}

To quantify the response as a function of time, we compute the
concentration correlation function for a given $x$-wavenumber $k_{x}$:
$\langle \tilde{n}(\bvec{x};k_x) \tilde{n}(\bvec{x}';k_x) \rangle = 2
\mbox{Re}\langle\hat{n}(y;k_x)\hat{n}^*(y';k_x)\rangle \cos
(k_x(x-x')) - 2 \mbox{Im}\langle\hat{n}(y;k_x)\hat{n}^*(y';k_x)\rangle
\sin (k_x(x-x'))$.  Fluctuations are strongly correlated near the
surfaces, even at equilibrium ($\we=0$).  At equilibrium, $\mbox{Im}
\langle \tilde{n}(y) \tilde{n}^*(y') \rangle$ is zero, but for nonzero
$\we$, it assumes values consistent with enhanced fluctuations whose
wave vectors are rotated from the flow direction into the first
quadrant, while the magnitude of the real part remains essentially
unchanged.  A snapshot of a typical concentration profile is shown in
Fig.~\ref{conc_noise} for $\we = 1, k_x = 1, b = 1$ and the
orientation, which is again consistent with the bulk HF mechanism, is
clear.  Hence, fluctuations form near the surface by mechanisms
analogous to those in the bulk, but are enhanced because of the
vicinity to the no-flux boundary.  Results for the no-slip case are
virtually identical -- the response of the flow to Brownian noise is
quite insensitive to the presence of slip.  Therefore, even in the
absence of flow instability, (i.e. when $\we\lesssim O(1)$ and/or slip
is absent), the near surface regions may make a nontrivial
contribution to the scattering signal, particularly its anisotropy
under flow.  The insensitivity to slip is consistent with what is
known about the eigenspectra of viscoelastic flows:
whether or not there is slip, the slowest decaying eigenmodes are
localized near solid boundaries.

These results clearly suggest that the bounding surfaces play an
important role in the formation of enhanced concentration fluctuations
in sheared polymer solutions.  In the context of the present model,
slip leads to a new class of viscoelastic flow instabilities which
result from the interaction of slip with stress and concentration, and
even in the absence of slip, Brownian fluctuations are selectively
enhanced near the surfaces.  These results are consistent with
experimental observations and underscore two points regarding the flow
behavior of polymeric liquids: (1) the distinctness and importance of
the dynamics of flowing polymers near boundaries, even at the
continuum level, and (2) the importance of couplings between various
phenomena for the dynamics and stability of these flows.

The authors gratefully acknowledge the National Science Foundation and
the Nontenured faculty grant program of the 3M Company for supporting
this work.  

\bibliographystyle{prsty}

\newpage

\noindent

\end{spacing}
\end{document}